\documentclass[aps,pre,twocolumn,groupedaddress,longbibliography]{revtex4-1}

\usepackage{graphicx}
\usepackage{bm}
\usepackage{amsmath}

\begin{document}

\title{Alignment of a topological defect by an activity gradient}
\author{Xingzhou Tang}
\author{Jonathan V. Selinger}
\affiliation{Department of Physics, Advanced Materials and Liquid Crystal Institute, Kent State University, Kent, Ohio 44242, USA}

\date{July 19, 2020}

\begin{abstract}
As a method for controlling active materials, researchers have suggested designing patterns of activity on a substrate, which should guide the motion of topological defects.  To investigate this concept, we model the behavior of a single defect of topological charge $+1/2$, moving in an activity gradient.  This modeling uses three methods:  (1) approximate analytic solution of hydrodynamic equations, (2) macroscopic, symmetry-based theory of the defect as an effective oriented particle, and (3) numerical simulation.  All three methods show that an activity gradient aligns the defect orientation, and hence should be useful to control defect motion.
\end{abstract}

\maketitle

\section{Introduction}

Topological defects are common in many areas of physics, from crystal structure to cosmology~\cite{Chaikin1995,Kleman2003}, and from bacterial growth to cell assembly~\cite{Duclos2017,Yaman2019}.  In conventional liquid crystals, topological defects are often used to identify phases, and the motion of topological defects is an important feature of coarsening dynamics.  In active liquid crystals, topological defects are continually forming in pairs, moving, and annihilating each other~\cite{Marchetti2013,Thampi2013,Giomi2013,Shi2013,Thampi2014,Guillamat2017,Lemma2019,Martinez-Prat2019}.  Theoretical research on two-dimensional (2D) active nematic liquid crystals has suggested that topological defects should be regarded as effective oriented particles~\cite{Vromans2016}.  In a previous article, we suggested that the defect orientation should be described by a tensor, with a tensor rank that depends on the topological charge of the defect~\cite{Tang2017}.  Further work has applied the concept of defects as effective oriented particles to defect motion induced by temperature, applied fields, fluid flow, boundary conditions, interactions with other defects, and activity~\cite{Shankar2018,Tang2019}.

Recently, one important research theme has been learning how to control active materials, in order to guide the flow of defects along pre-selected paths.  Two articles have proposed to achieve this control by designing patterns of activity on a substrate.  In that way, the activity gradient aligns the orientations of the defects with topological charge of $+1/2$, through a mechanism analogous to an electric field aligning electric dipole moments~\cite{Shankar2019,Zhang2019}.  Hence, the patterns of activity create paths for the motion of defects.

In this article, we further investigate the concept of topological defects moving in a system with nonuniform activity.  Rather than studying the system as a whole, we concentrate on a single defect of topological charge $+1/2$ moving in an activity gradient, and analyze the dynamics using three complementary approaches.  First, we use an approximate analytic method to solve the hydrodynamic equations for the liquid-crystal director and the flow field around a defect.  Second, we construct a macroscopic symmetry-based theory for the defect as an oriented particle, and use it to determine the interaction with an activity gradient.  Third, we perform finite-element simulations of the defect motion, and determine how the orientation evolves in response to the activity gradient.  We show that all three approaches give consistent descriptions of defect alignment, and support the concept of controlling active materials with patterns of activity.

\section{Hydrodynamic Theory}

As a first step, we consider the hydrodynamic equations around a $+1/2$ defect.  We follow the method of our previous article~\cite{Tang2019}, but now add a gradient of the activity.  For this calculation, the nematic order is described by the director field $\hat{\bm{n}}(\bm{r},t)=(\cos\theta(\bm{r},t),\sin\theta(\bm{r},t))$.  With the approximation of equal Frank elastic constants, the Frank free energy is
\begin{equation}
F=\int d^2 r \left[\frac{1}{2}K(\partial_i n_j)(\partial_i n_j)\right]=\int d^2 r \left[\frac{1}{2}K |\bm{\nabla}\theta|^2\right].
\end{equation}
The dynamic evolution of the director $\hat{\bm{n}}$ is coupled with the fluid flow velocity field $\bm{v}(\bm{r},t)$.  There are two modes that dissipate energy:  the strain rate tensor $A_{ij}=\frac{1}{2}(\partial_i v_j + \partial_j v_i)$ and the director rotation with respect to the background fluid vorticity $N_i = \dot{n}_i - \omega\epsilon_{ji}n_j$, where $\epsilon_{ji}$ is the 2D Levi-Civita symbol and $\omega=\frac{1}{2}\epsilon_{km}\partial_k v_m$.  In terms of these modes, a minimal model for the dissipation function is
\begin{equation}
D=\int d^2 r \left[\frac{1}{2}\alpha_4 A_{ij}A_{ij}+\frac{1}{2}\gamma_1 N_i N_i - 2\zeta(\bm{r})n_i n_j A_{ij}\right].
\end{equation}
Here, the first term represents the dissipation from conventional fluid flow, and the second term represents dissipation from rotation of the nematic order with respect to the fluid.  The third term is an extra contribution arising from activity $\zeta(\bm{r})$, which we allow to be nonuniform.  This term is really a ``rate of energy input'' (with a negative sign) rather than an ``energy dissipation,'' but it still enters into the theoretical formalism of a dissipation function.  The sign $\zeta>0$ corresponds to extensile activity, and $\zeta<0$ to contractile activity.

We assume that the material is incompressible, which implies that $\partial_i v_i = 0$.  Because of this constraint, the velocity field can be written in terms of a stream function $\psi(\bm{r},t)$ as $v_i = \epsilon_{ij}\partial_j \psi$.

From the free energy and the dissipation function, we derive the equations of motion for $\theta$ and $\psi$.  For the director orientation $\theta$, the elastic force is $-\delta F/\delta\theta(\bm{r},t)$, and the drag force is $-\delta D/\delta[\partial_t\theta(\bm{r},t)]$.  Hence, the equation for overdamped motion is that the forces sum to zero,
\begin{equation}
-\frac{\delta F}{\delta\theta(\bm{r},t)}-\frac{\delta D}{\delta[\partial_t\theta(\bm{r},t)]}=0.
\end{equation}
For the generalized velocity $\psi$, the elastic force is zero, and the combined drag and active force is $-\delta D/\delta\psi(\bm{r},t)$.  Hence, the equation for overdamped motion is that this combined drag and active force equals zero,
\begin{equation}
-\frac{\delta D}{\delta\psi(\bm{r},t)}=0.
\end{equation}
The detailed expressions for these equations are worked out in Ref.~\cite{Tang2019}.

We now apply these general equations to the specific case of nonuniform activity of the form $\zeta(\bm{r})=\zeta' y$, so that the activity gradient is $\bm{\nabla}\zeta=\zeta'\hat{\bm{y}}$.  We assume that the activity gradient is small, so that we can use perturbation theory in $\zeta'$, following the approach of Ref.~\cite{Pismen1990}.  For this perturbation theory, we write the steady-state solution as
\begin{subequations}
\begin{align}
\label{thetaseries}
\theta(\bm{r})&=\theta_0(\bm{r})+\zeta'\theta_1(\bm{r})+O(\zeta'^2)\\
\psi(\bm{r})&=\psi_0(\bm{r})+\zeta'\psi_1(\bm{r})+O(\zeta'^2).
\end{align}
\end{subequations}

At zero-th order in $\zeta'$, the stream function $\psi_0(\bm{r})$ is constant, which implies that the flow velocity is zero.  The differential equation for $\theta_0$ then becomes Laplace's equation $\nabla^2 \theta = 0$.  A solution corresponding to a $+1/2$ defect at the origin is $\theta_0(\bm{r})=\frac{1}{2}\tan^{-1}(y/x)+\Theta_0$.  From Ref.~\cite{Tang2017}, the defect orientation is the orientation in which the director points outward from the defect.  It is given by the vector $\bm{p}=(\cos\Psi,\sin\Psi)$, with $\Psi=2\Theta_0$.

We now go to first order in perturbation theory in $\zeta'$.  We insert the zero-th order expressions for $\theta_0$ and $\psi_0$ into the partial differential equations, and solve for the first-order corrections $\theta_1$ and $\psi_1$.  We calculate these solutions in polar coordinates with boundary conditions such that the first-order correction does not change the far-field behavior, hence $\theta_1(r_\text{max},\phi)=0$, $\psi_1(r_\text{max},\phi)=0$, and $\partial_r\psi_1(r_\text{max},\phi)=0$.  We put $\theta_0$ and $\theta_1$ into the series of Eq.~(\ref{thetaseries}), and put that series into the director field $\bm{n}$.  We then determine the perturbed orientation in which the director points outward from the defect,
\begin{equation}
\Psi=2\Theta_0+\frac{\zeta' r_\text{max}^3\cos(2\Theta_0)}{6 K}\frac{g(4+2g-\sqrt{4+2g})}{(3+2g)(2+g+\sqrt{4+2g})},
\label{Psiprediction}
\end{equation}
where $g=\gamma_1/\alpha_4$.  Hence, we can see that the activity gradient $\zeta'$ generates a correction term, which shifts the defect orientation.

To interpret the correction term, suppose that $\zeta'>0$.  If $\cos(2\Theta_0)>0$, so that the unperturbed $\bm{p}$ has a positive $x$ component, then the correction term is positive.  It rotates the defect orientation counter-clockwise toward the positive $y$ axis.  If $\cos(2\Theta_0)<0$, so that the unperturbed $\bm{p}$ has a negative $x$ component, then the correction term is negative, and it rotates the defect orientation clockwise toward the positive $y$ axis.  In both cases, the correction is toward the positive $y$ axis, so toward the $\bm{\nabla}\zeta$ direction.  Likewise, if $\zeta'<0$, the correction is toward the negative $y$ axis, which is then toward the $\bm{\nabla}\zeta$ direction.  Thus, in general, the activity gradient tends to reorient the $+1/2$ defect orientation toward the activity gradient direction.  This effect is analogous to an electric field acting on an electric dipole moment, which tends to reorient the dipole toward the electric field direction.

\section{Macroscopic Theory}

As an alternative approach, we consider the same problem of defect alignment from the perspective of a macroscopic, symmetry-based theory.

In Ref.~\cite{Tang2019}, we argued that defect motion can be described in terms of the macroscopic degrees of freedom for the defect, without considering the director or the flow velocity field around the defect.  In this perspective, the defect is an oriented particle, with a position $\bm{R}(t)$ and some further variable to represent the orientation.  For a defect of topological charge $+1/2$, the orientation is represented by a unit vector $\bm{p}(t)=(\cos\Psi(t),\sin\Psi(t))$.  We can then construct the free energy and the dissipation function in terms of those macroscopic variables, using the most general forms allowed by symmetry.

In a conventional, non-active liquid crystal, the dissipation function should be a quadratic form in the defect velocity $\dot{\bm{R}}$ and the rotational velocity $\dot{\bm{p}}$.  Hence, the most general form for the dissipation function is~\cite{Tang2019}
\begin{equation}
D_\text{passive} = \frac{1}{2}D_1 |\dot{\bm{R}}|^2 + \frac{1}{2}D_2 (\bm{p}\cdot\dot{\bm{R}})^2 + \frac{1}{2}D_3 |\dot{\bm{p}}|^2
+ D_4 \dot{\bm{p}}\cdot\dot{\bm{R}} .
\end{equation}
Here, the $D_1$ term gives the energy dissipated by defect translation, and the $D_2$ term shows how that dissipation depends on the defect orientation with respect to the velocity.  The $D_3$ term gives the energy dissipated by defect rotation, and the $D_4$ term shows a dissipative coupling between translation and rotation.  The quadratic form is positive-definite if $D_4^2 < D_1 D_3$.

In an active liquid crystal, the dissipation function is not required to be quadratic in $\dot{\bm{R}}$ and $\dot{\bm{p}}$.  Rather, it may have active terms, which are linear in $\dot{\bm{R}}$ or $\dot{\bm{p}}$, and hence break time-reversal symmetry.  For an active liquid crystal with \emph{uniform} activity, there is one active term,
\begin{equation}
D_\text{active}^\text{uniform}=d_5 \zeta \bm{p}\cdot\dot{\bm{R}}.
\end{equation}
(In Ref.~\cite{Tang2019}, we wrote the coefficient as $D_5=d_5 \zeta$, but here we explicitly show the dependence on the activity parameter $\zeta$.)  In the uniform system, there is no active term involving $\dot{\bm{p}}$, because we cannot construct a scalar that is linear in $\dot{\bm{p}}$.  The combination $\bm{R}\cdot\dot{\bm{p}}$ is forbidden by translational invariance, the combination $\dot{\bm{R}}\cdot\dot{\bm{p}}$ is a non-active, quadratic term, and the combination $\bm{p}\cdot\dot{\bm{p}}=0$ because $\bm{p}$ is a unit vector.

Now consider an active liquid crystal with \emph{nonuniform} activity.  In that case, it is possible to construct an active term involving $\dot{\bm{p}}$,
\begin{equation}
D_\text{active}^\text{nonuniform}=d_6 (\bm{\nabla}\zeta)\cdot\dot{\bm{p}}.
\label{Dactivenonuniform}
\end{equation}
In this perspective, the activity gradient $\bm{\nabla}\zeta$ is important because it creates an extra vector that can couple to $\dot{\bm{p}}$.  The total dissipation function is then $D_\text{total}=D_\text{passive}+D_\text{active}^\text{uniform}+D_\text{active}^\text{nonuniform}$.

In a recent article~\cite{Tang2020}, we argued that a dissipation function can generate an effective potential, which induces steady-state alignment.  Here, let us apply that general argument to $D_\text{active}^\text{nonuniform}$.  This term can be rewritten using the defect orientation angle $\Psi$ as 
\begin{equation}
D_\text{active}^\text{nonuniform}=d_6 [-(\partial_x\zeta)\sin\Psi+(\partial_y\zeta)\cos\Psi]\dot{\Psi}.
\end{equation}
It generates a force acting on $\Psi$,
\begin{align}
f_\text{active}^\text{nonuniform}&=-\frac{\partial D_\text{active}^\text{nonuniform}}{\partial\dot{\Psi}}\\
&=-d_6 [-(\partial_x\zeta)\sin\Psi+(\partial_y\zeta)\cos\Psi].\nonumber
\end{align}
That force generates an effective potential acting on $\Psi$ in the steady state,
\begin{align}
U_\text{active}^\text{nonuniform}&=-\int f_\text{active}^\text{nonuniform}d\Psi\nonumber\\
&=d_6 [(\partial_x\zeta)\cos\Psi+(\partial_y\zeta)\sin\Psi]\nonumber\\
&=d_6 (\bm{\nabla}\zeta)\cdot\bm{p}.
\end{align}
This effective potential has the same mathematical form as an electric field interacting with an electric dipole moment.  It tends to align the defect orientation vector $\bm{p}$ along the activity gradient $\bm{\nabla}\zeta$.

This macroscopic, symmetry-based theory has both disadvantages and advantages compared with the hydrodynamic theory of the previous section.  One disadvantage is that the macroscopic theory does not tell us whether the coefficient $d_6$ is positive or negative, i.e.\ whether the alignment of $\bm{p}$ is parallel or antiparallel to $\bm{\nabla}\zeta$.  Based on the hydrodynamic theory, we can assume that the favored alignment of $\bm{p}$ is parallel to $\bm{\nabla}\zeta$, so that $d_6<0$.

By comparison, one advantage of the macroscopic theory is that it can easily be applied to defects of different symmetries.  For example, consider a defect of topological charge $-1/2$.  Because this defect has three-fold symmetry, its orientation can be described by a third-rank, completely symmetric tensor $T_{ijk}$, as discussed in Ref.~\cite{Tang2017}.  The macroscopic theory shows us that there is no linear coupling between the orientation tensor $T_{ijk}$ and the activity gradient $\bm{\nabla}\zeta$.  Rather, there may be a higher-order coupling of the form $(\partial_i\zeta)(\partial_j\zeta)(\partial_k\zeta)T_{ijk}$, or a coupling with a third derivative of the form $(\partial_i\partial_j\partial_k\zeta)T_{ijk}$.  Hence, even without any calculations, we can see that a $-1/2$ defect would not be as strongly aligned by an activity gradient as a $+1/2$ defect, but it would have these weaker, higher-order aligning interactions.

\section{Simulations}

So far, we have used two types of analytic theory to show that the orientation of a $+1/2$ defect is aligned by an activity gradient.  As a numerical test of these analytic arguments, we now perform simulations of the dynamic evolution of the position and orientation of a $+1/2$ defect in an activity gradient.

For the simulations, we follow the method of Ref.~\cite{Tang2019}.  In this method, we allow both the magnitude and the direction of nematic order to vary, so that defects will be able to form and move freely.  Hence, we represent nematic order by the tensor $Q_{ij}(\bm{r},t)=S(\bm{r},t)[2n_i(\bm{r},t)n_j(\bm{r},t)-\delta_{ij}]$, with magnitude $S$ and director $\hat{\bm{n}}$.  The magnitude $S$ goes to zero at the defect core.  With the approximation of equal Frank constants, the Landau-de Gennes free energy for this model is
\begin{align}
F=\int d^2 r \biggl[&-\frac{1}{4}a Q_{ij}Q_{ij}+\frac{1}{16}b(Q_{ij}Q_{ij})^2\nonumber\\
&+\frac{1}{16}L(\partial_k Q_{ij})(\partial_k Q_{ij})\biggr].
\end{align}
This free energy favors a bulk order parameter $S=(a/b)^{1/2}$ away from any defect, and it favors a defect core radius $r_\text{core}=(L/a)^{1/2}$.  The elastic constant $L$ in this tensor representation is related to the Frank constant $K$ in the director representation by $K=L S^2$.

In this tensor representation, the two modes that dissipate energy are the strain rate tensor $A_{ij}=\frac{1}{2}(\partial_i v_j + \partial_j v_i)$ and the rotation of nematic order with respect to the background fluid vorticity $B_{ij}=\dot{Q}_{ij}-{\omega(\epsilon_{lj}Q_{il}+\epsilon_{li}Q_{lj})}$, where $\epsilon_{ji}$ is the 2D Levi-Civita symbol and $\omega=\frac{1}{2}\epsilon_{km}\partial_k v_m$.  In terms of these modes, a minimal model for the dissipation function is
\begin{equation}
D=\int d^2 r \biggl[\frac{1}{2}\alpha_4 A_{ij}A_{ij}+\frac{1}{16}\Gamma_1 B_{ij}B_{ij}-Z(\bm{r})Q_{ij}A_{ij}\biggr].
\end{equation}
Here, the first term represents the dissipation from conventional fluid flow, the second term represents rotation of nematic order with respect to the fluid, and the third term arises from the activity $Z(\bm{r})$, which may be nonuniform.  The rotational viscosity and activity coefficients in the tensor representation are related to the corresponding coefficients in the director representation by $\gamma_1 = \Gamma_1 S^2$ and $\zeta = Z S$.

Based on the free energy and dissipation function, the partial differential equation for overdamped dynamics of the nematic order tensor becomes
\begin{equation}
0=-\frac{\delta F}{\delta Q_{ij}}-\frac{\delta D}{\delta\dot{Q}_{ij}}.
\end{equation}
Similarly, the equation for inertial dynamics of the flow velocity field becomes
\begin{equation}
\rho(\partial_t + v_j \partial_j)v_i = -\frac{\delta D}{\delta v_i} + \partial_i p,
\end{equation}
where $\rho$ is the mass density and $p(\bm{r},t)$ is the pressure.  Finally, the constraint of incompressibility is $\partial_i v_i = 0$.

We solve these equations numerically with the finite-element method, using the software package COMSOL, inside a circular domain of radius $r_\text{max}$.  For the initial condition, we impose a $+1/2$ defect at the center, with its initial orientation in the $+x$ direction.  At later times, we find the location of the defect by searching for the minimum of the scalar order parameter $S(\bm{r},t)=[\frac{1}{2}Q_{ij}Q_{ij}]^{1/2}$.  After finding the defect, we determine the defect orientation vector $\bm{p}=(\bm{\nabla}\cdot\bm{Q})/|\bm{\nabla}\cdot\bm{Q}|$, and hence the angle $\Psi=\tan^{-1}(p_y/p_x)$.

We use parameters $a=b=200$, $L=4$, $\alpha_4=1$, $\Gamma_1=8$, and $\rho=1$ (in arbitrary units).  With these parameters, the bulk order parameter is $S=1$, and the defect core radius is $r_\text{core}\approx0.2$.  We impose a nonuniform activity pattern $Z(x,y)=Z' y$, with a gradient in the $+y$ direction.  We vary the activity gradient $Z'$ and the system size $r_\text{max}$, as discussed below.

For a first set of simulations, we apply Dirichlet boundary conditions at $r_\text{max}$, so that the nematic order at the boundary is fixed in its initial configuration.  As we move forward in time, the defect position shifts and its orientation rotates.  However, the system is highly constrained because of the boundary condition.  Because of this constraint, the system reaches a steady state, with a limited translation and rotation of the defect.

\begin{figure}
\includegraphics[width=\columnwidth]{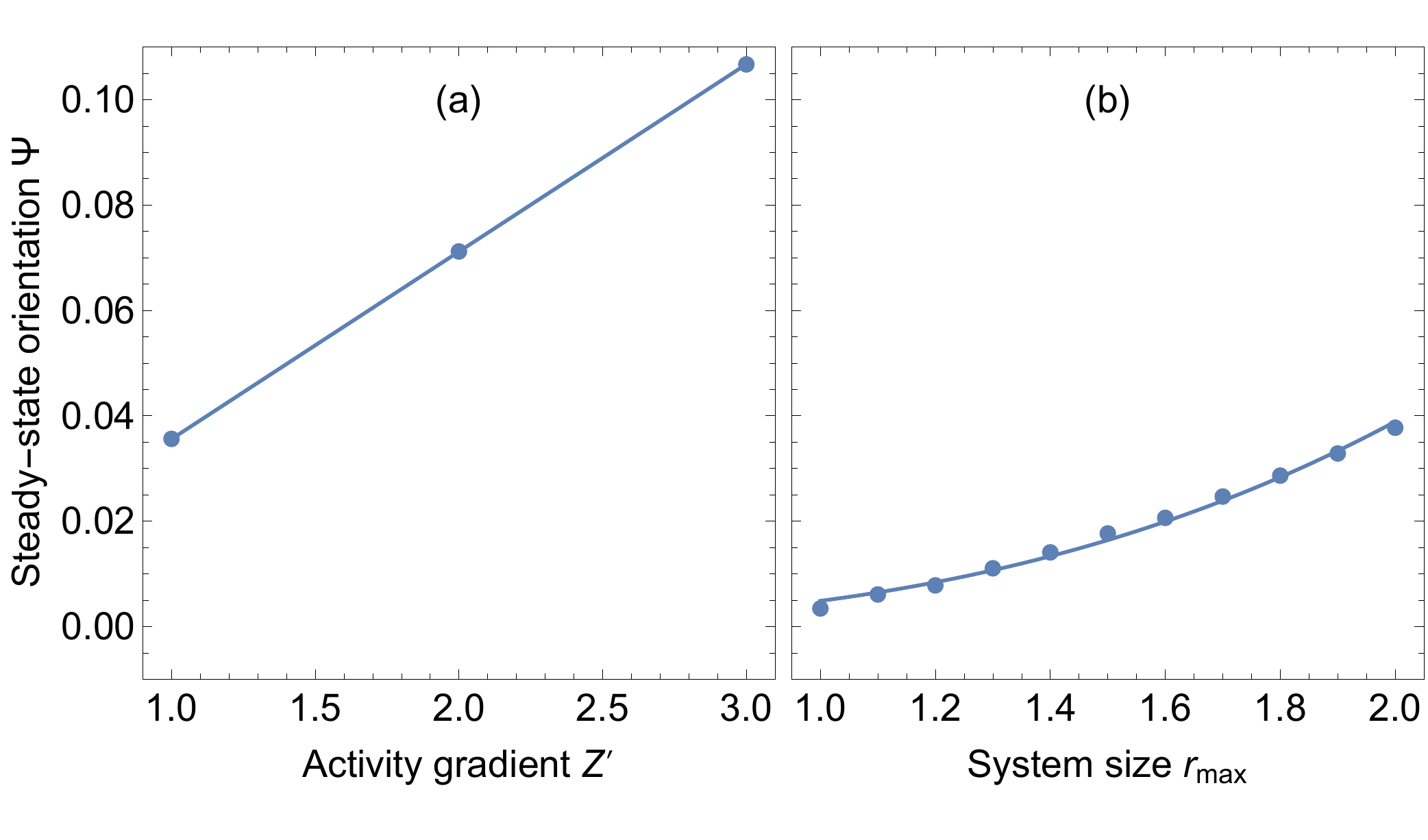}
\caption{(a)~Steady-state defect orientation $\Psi$ as a function of activity gradient $Z'$, at $r_\text{max}=2$, compared with a linear fit.  (b)~Steady-state defect orientation $\Psi$ as a function of system size $r_\text{max}$, at fixed $Z'=1$, compared with a cubic fit.}
\end{figure}

Figure~1(a) shows the numerical results for the steady-state defect orientation angle $\Psi$ as a function of $Z'$, with fixed $r_\text{max}=2$.  We can see that $\Psi$ is linearly proportional to $Z'$.  Similarly, Fig.~1(b) shows corresponding results for $\Psi$ as a function of $r_\text{max}$ for fixed $Z'=1$.  Here, we see that $\Psi$ is proportional to $r_\text{max}^3$, because the larger system size allows the defect more freedom to rotate.  Both of these results are consistent with the scaling predicted in Eq.~(\ref{Psiprediction}).

\begin{figure}
\includegraphics[height=\columnwidth]{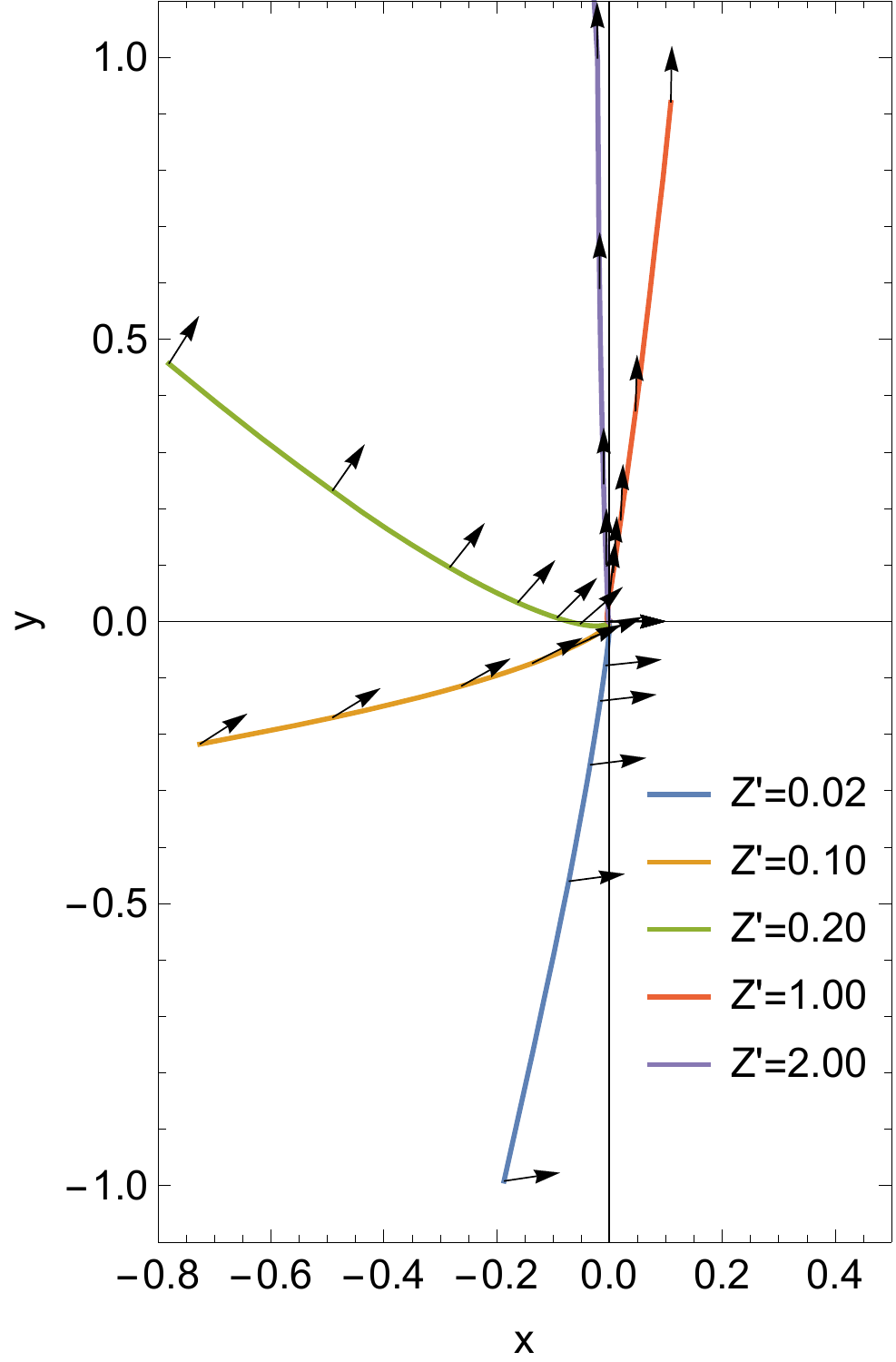}
\caption{Examples of defect trajectories, as a defect moves outward from the origin and exits the system with free boundary conditions.  The arrows indicate the defect position and orientation at times $t=0$, 2.0, 2.5, 3.0, 3.5, 4.0, \ldots.}
\end{figure}

\begin{figure}
\includegraphics[width=\columnwidth]{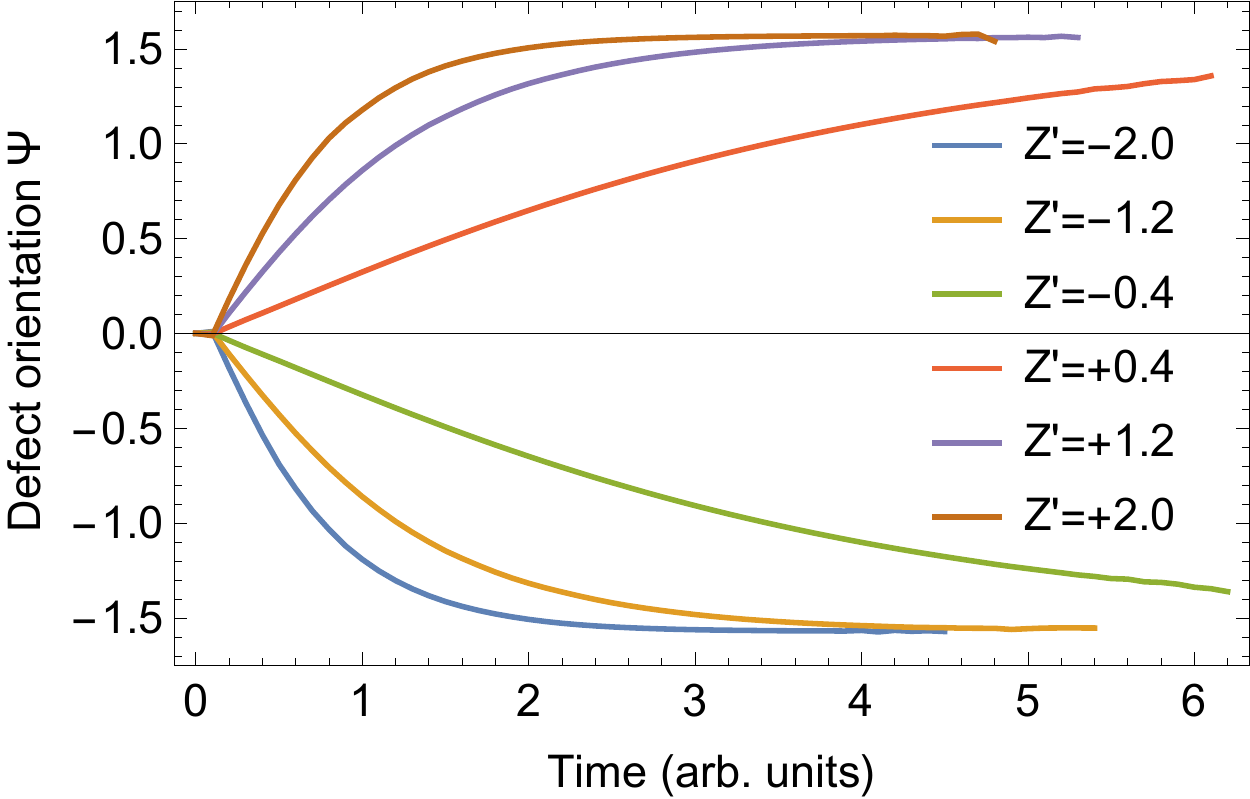}
\caption{Time evolution of the defect orientation angle $\Psi$, in a system with free boundary conditions, for several positive and negative values of the activity gradient $Z'$.}
\end{figure}

For another set of simulations, we apply free boundary conditions, so that the defect is not constrained by the boundaries.  With these boundary conditions, the defect will eventually move out of the system, but we can monitor its position and orientation until it leaves.  Figure~2 shows examples of the defect trajectories, with arrows indicating the defect positions and orientations at the specified times.  We can see that the orientation rotates toward alignment with the activity gradient in the $+y$ direction.  This rotation occurs quickly for large $Z'$, and more slowly for small $Z'$, so that the rotation may not be complete before the defect exits the system.

Similarly, Fig.~3 shows the time evolution of the defect orientation angle $\Psi$ in the system with free boundary conditions, with several positive and negative values of the activity gradient $Z'$.  These results show that $\Psi$ is driven toward $+\pi/2$ when $Z'>0$, and it is driven toward $-\pi/2$ when $Z'<0$.  In both cases, the defect orientation vector $\bm{p}=(\cos\Psi,\sin\Psi)$ is driven to be parallel to the activity gradient $\bm{\nabla}Z=Z'\hat{\bm{y}}$.

For a final example, we consider a more complex pattern of activity.  In this example, we are inspired by a recent experiment~\cite{norton2018insensitivity}, which investigated circulating motion of two $+1/2$ defects in a circular disk with uniform activity.  For a related nonuniform simulation, we construct a system with a circular region of nonzero activity in the center, surrounded by a larger region of zero activity.  We expect that the activity gradient between the two regions will confine defects, so that they will circulate around this effective confined region.

\begin{figure}
	\centering
	\begin{tabular*}{0.5\textwidth}{@{\extracolsep{\fill}}cc}
		\includegraphics[width=4.5cm]{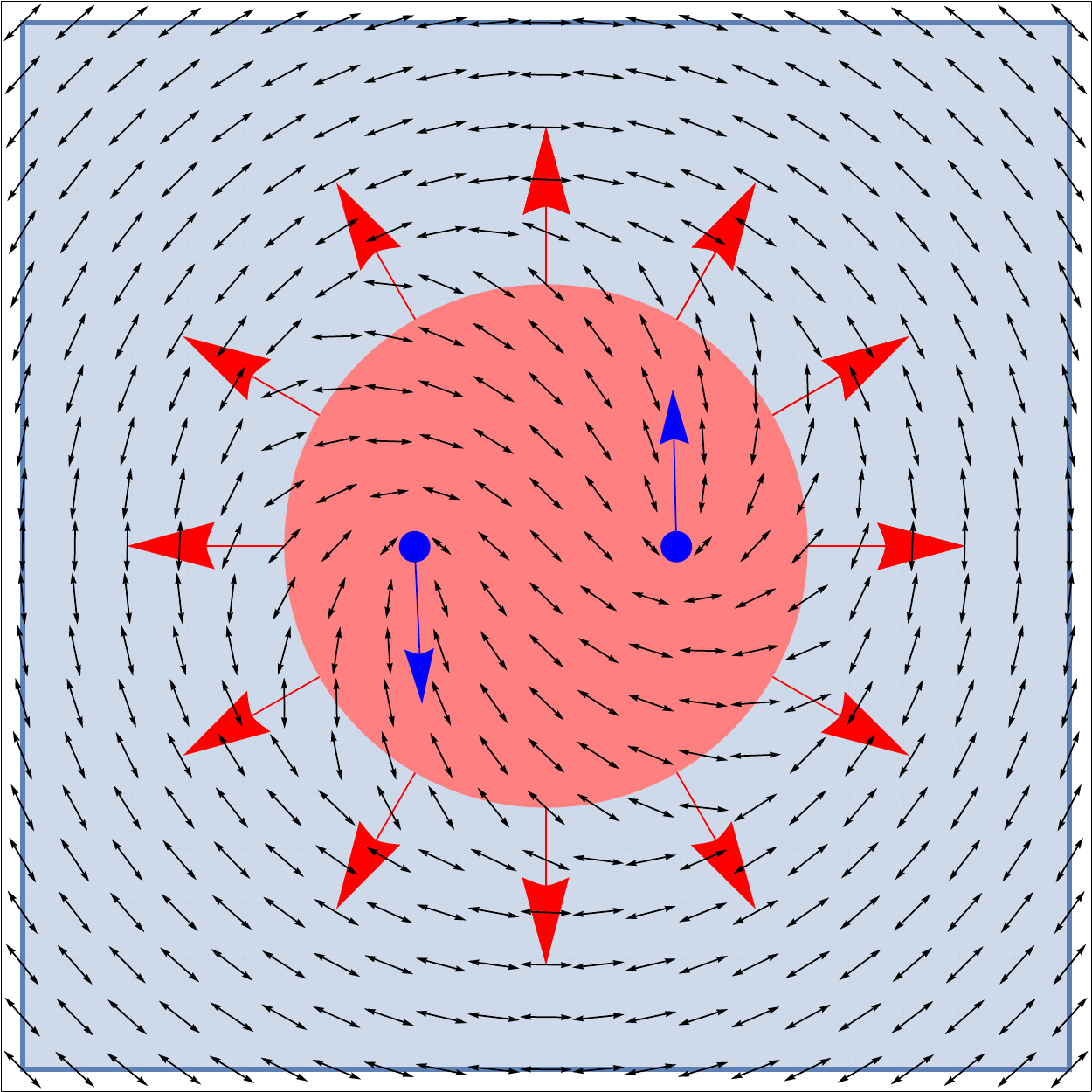} & \includegraphics[width=4.5cm]{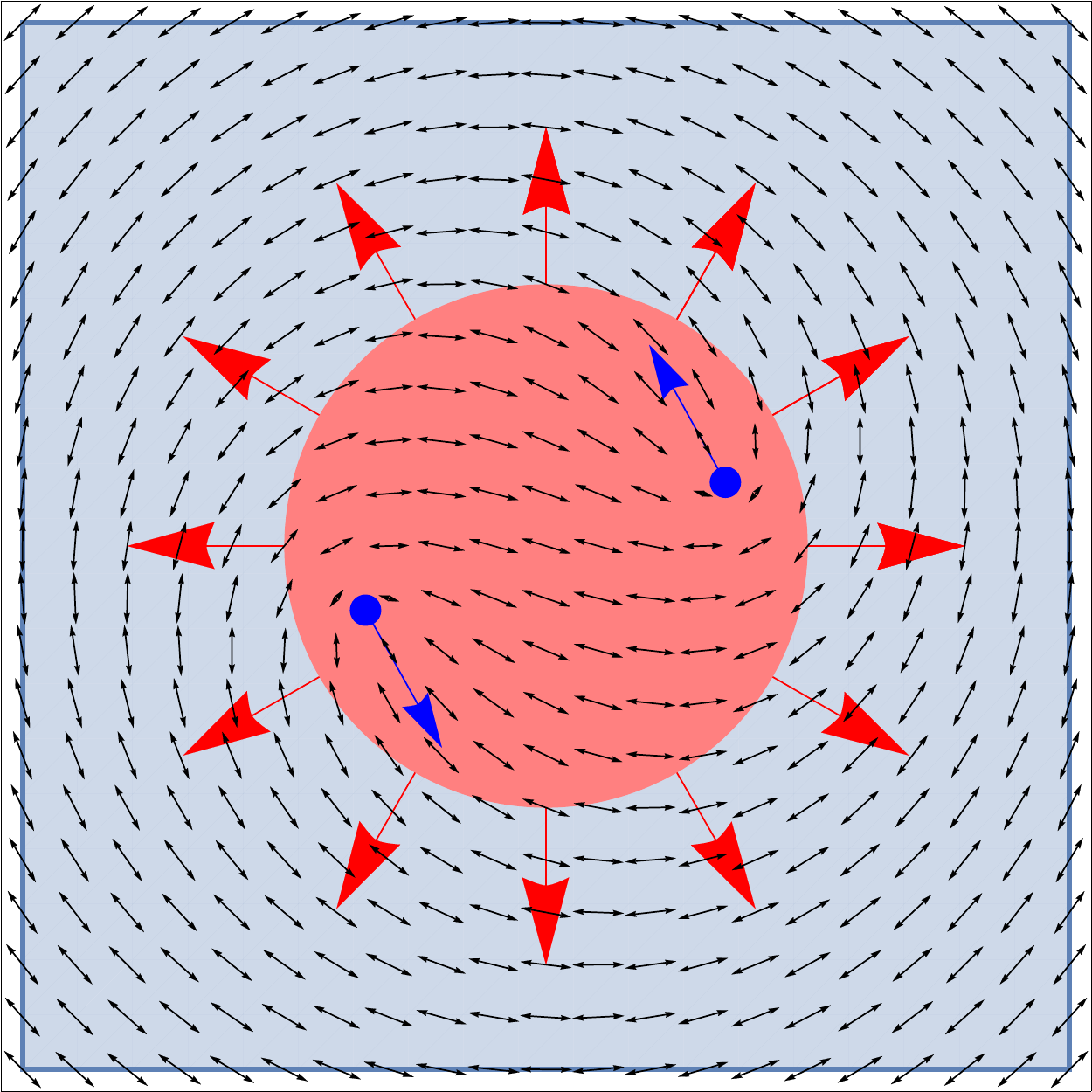} \\
		(a) & (b)\\
		\includegraphics[width=4.5cm]{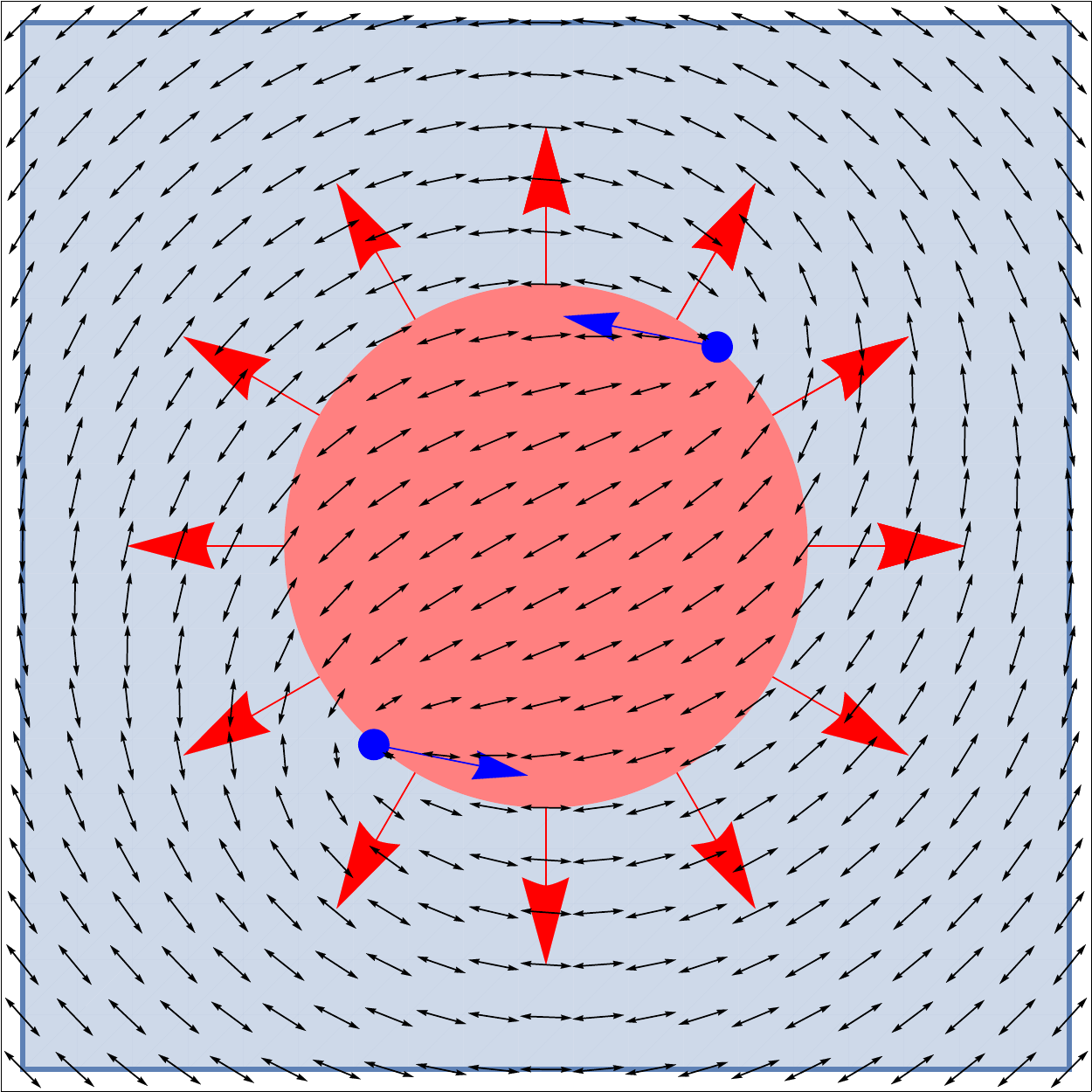} & 
		\includegraphics[width=4.5cm]{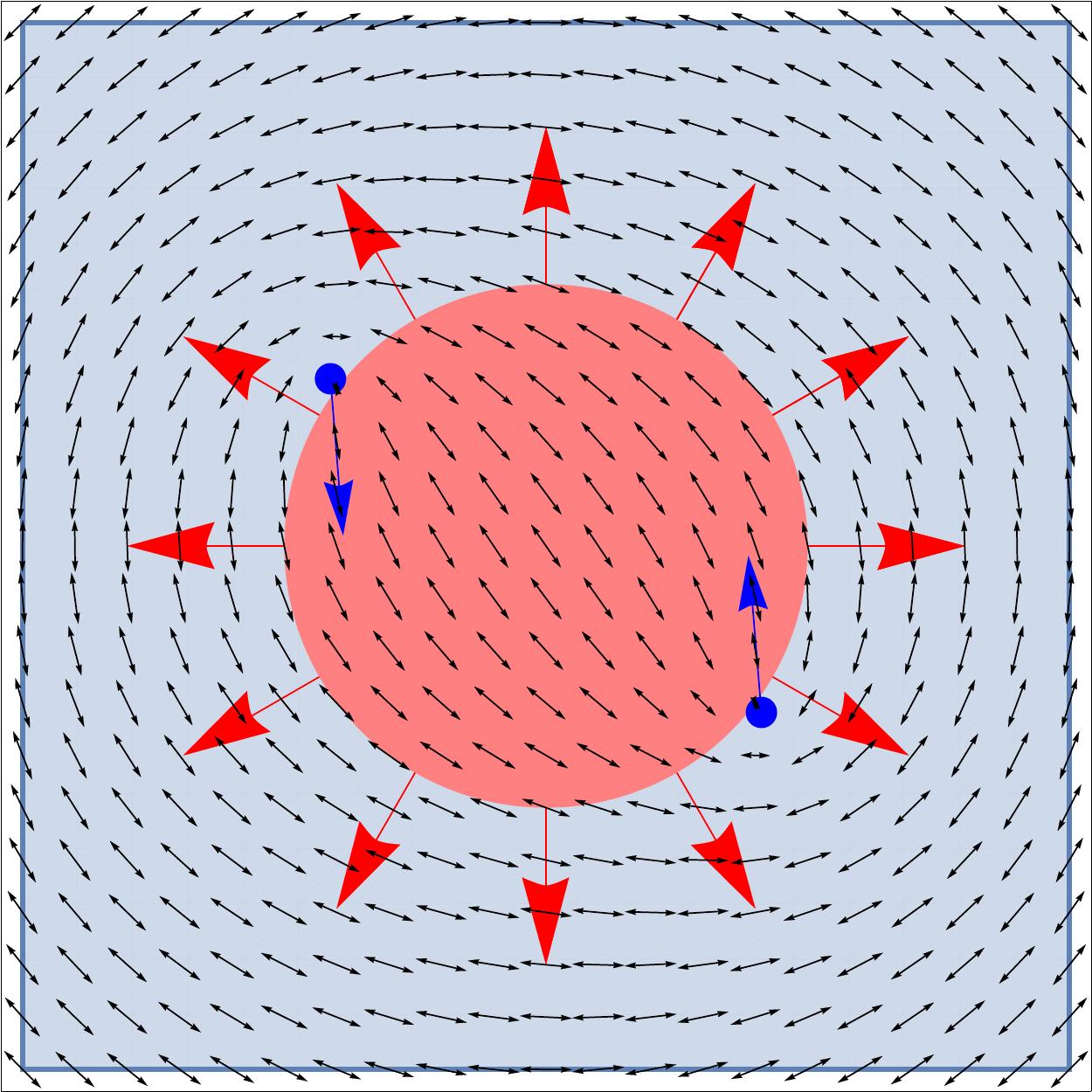} \\
		(c) & (d)\\
		\includegraphics[width=4.5cm]{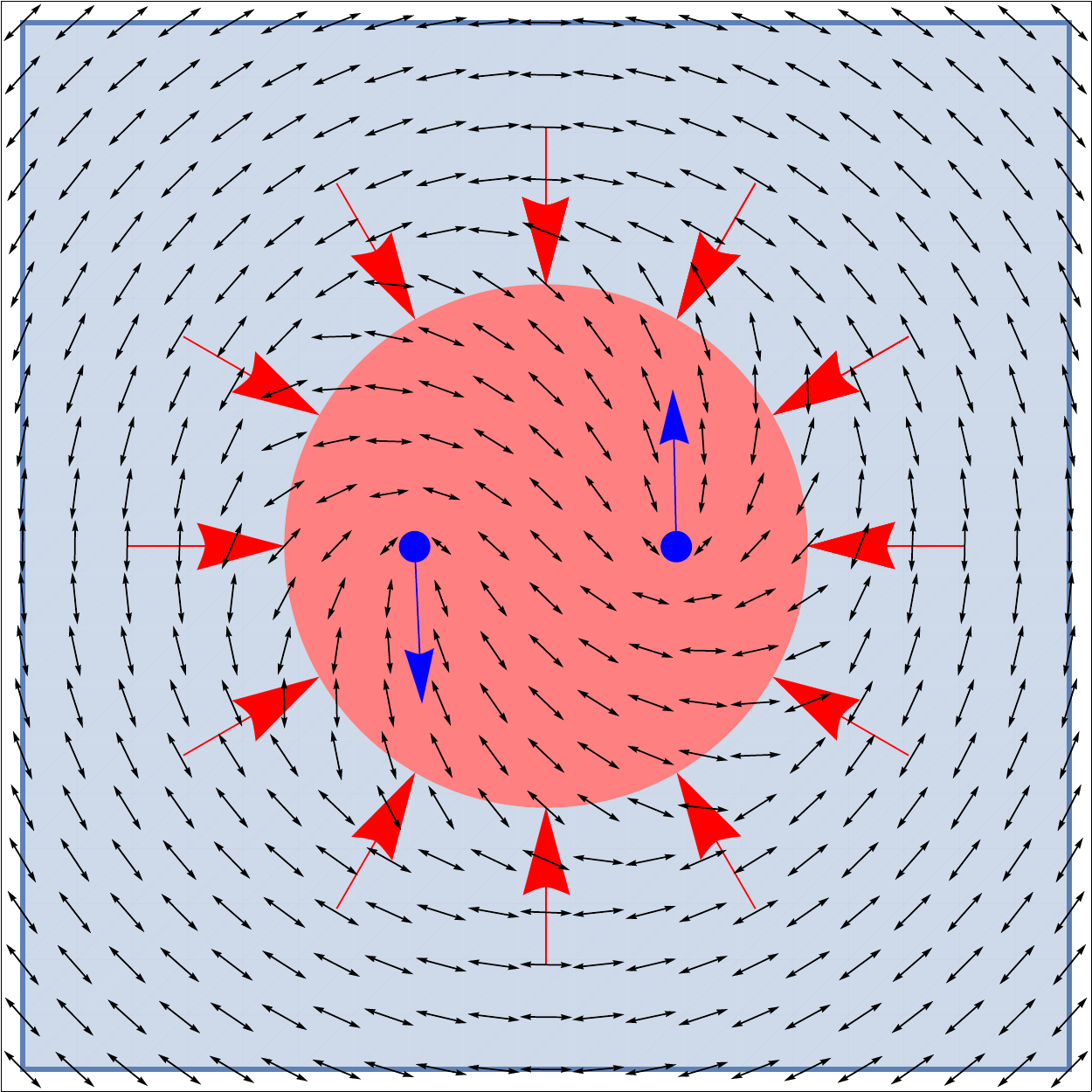} & \includegraphics[width=4.5cm]{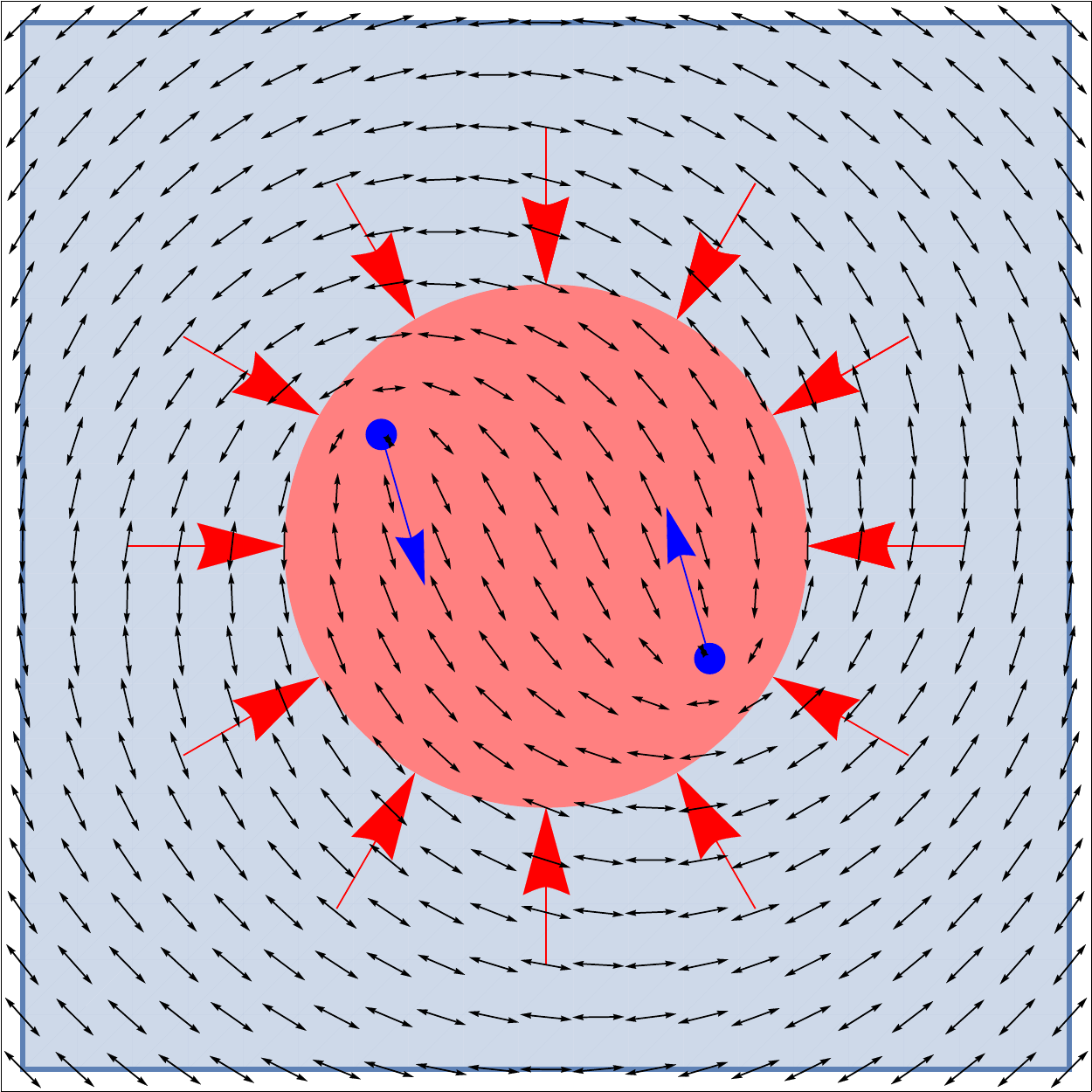} \\
		(e) & (f)\\
		\includegraphics[width=4.5cm]{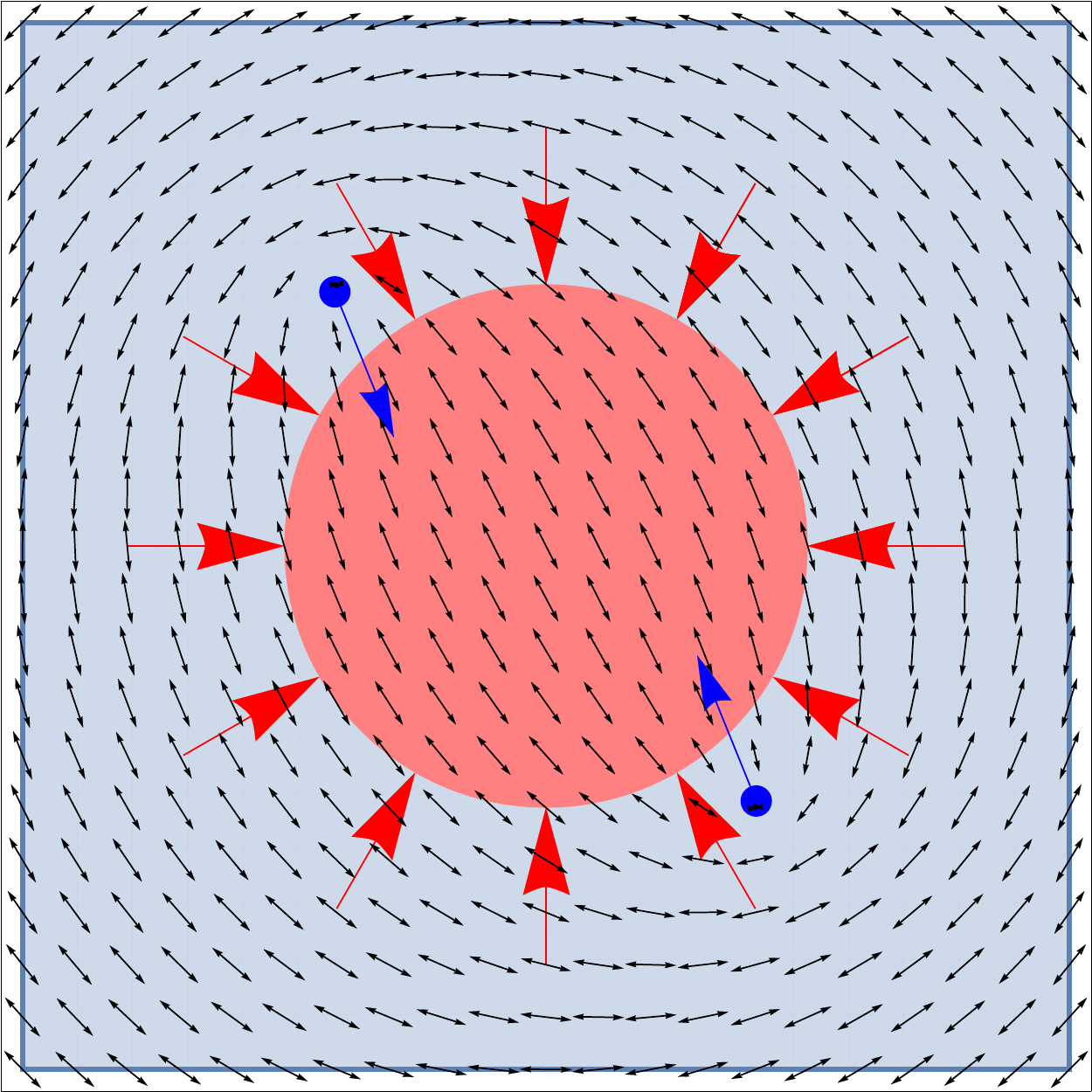} & 
		\includegraphics[width=4.5cm]{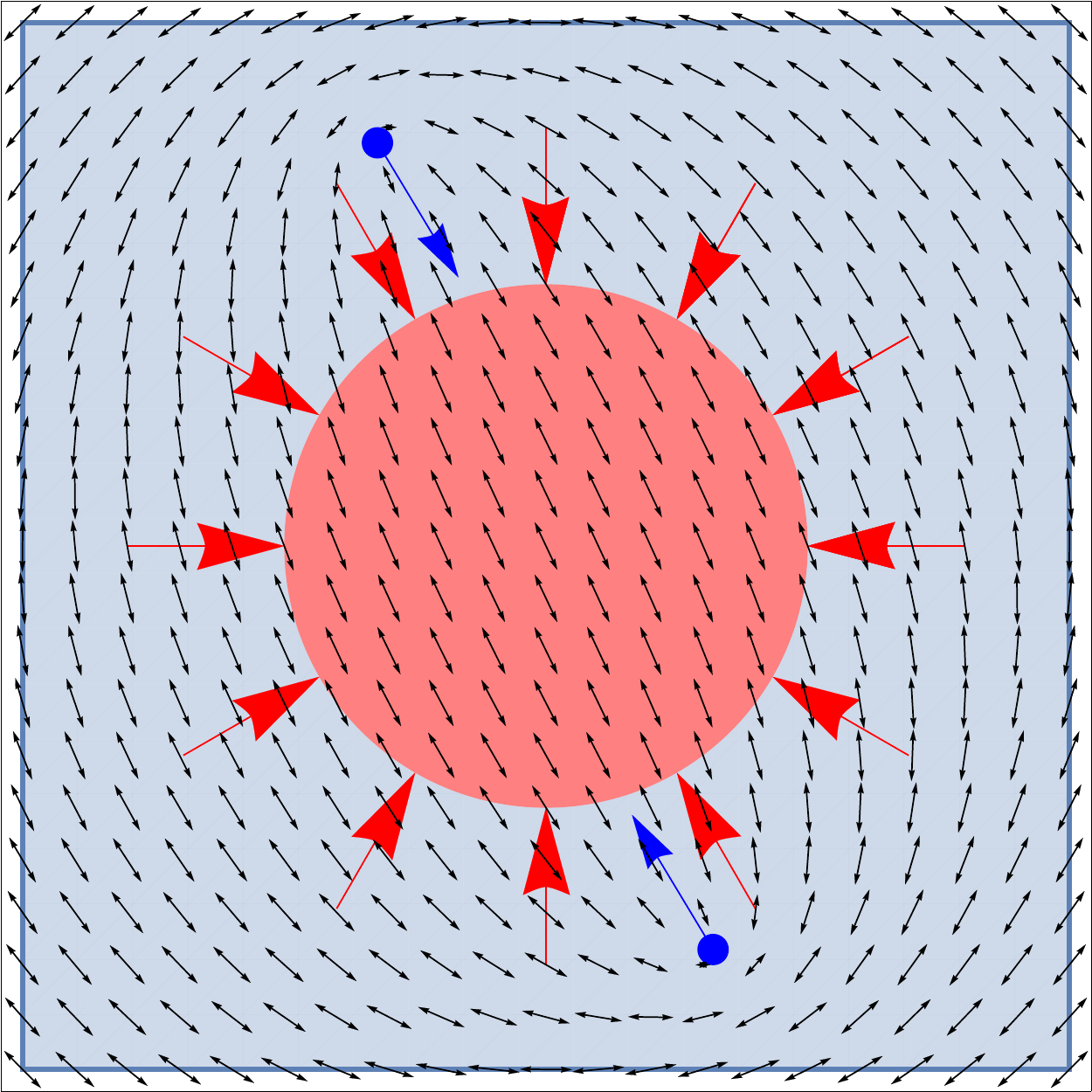} \\
		(g) & (h)
		\label{circularmotion}
	\end{tabular*}
	\caption{Snapshots of the dynamic evolution of the circular motion of $+1/2$ defects. Figure (a) to (d) shows the process with negative activity, and Figure (e) to (f) shows the process with positive activity. The pink circle is the area with activity, and the red arrows show the direction of activity gradient. The blue arrows show the location and orientation of defects.}
\end{figure}

The simulation geometry is shown in Fig.~4. The pink circle in the center is the active region, while the grey region around the circle is not active.  The activity profile is
\begin{equation}
Z(x,y) = -\frac{Z_0}{\pi}\tan^{-1} \left[C \left(x^2+y^2-r_0^2 \right) \right]+\frac{1}{2},
\end{equation}
where $Z_0$ is the value of activity within the circular region, $r_0$ is the radius of the region, and $C$ is a coefficient to determine the steepness of the activity gradient.  If $Z_0<0$, the gradient points outward, as shown by the red arrows in Fig.~4(a-d).  Likewise, if $Z_0>0$, the gradient points inward, shown by the red arrows in Fig.~4(e-h).

The boundary condition on the square outer edge of the grey region is tangential, with $\theta_{BC}(x,y)=\tan^{-1}(y/x)+\pi/2$.  This boundary condition requires a total topological charge of $+1$ inside the system, presumably in the form of two defects of topological charge $+1/2$ each.  Furthermore, this boundary condition favors an \emph{inward} alignment of the $+1/2$ defect orientation vectors, shown by the blue arrows on the defects.

We set an initial state with a pair of $+1/2$ defects in the active region, one pointing upward and the other downward. Inside the active region, the initial director field is
\begin{equation} 
\theta(x,y)=\frac{1}{2}\left[\tan^{-1}\left(\frac{y-y_1}{x- x_1}\right)+\tan^{-1}\left(\frac{y- y_2}{x- x_2}\right)\right]+\frac{\pi}{4},
\end{equation} 
where $(x_1,y_1)$ and $(x_2, y_2)$ are the locations of these two $+1/2$ defects.  Outside the active region, the initial director field just matches the boundary condition.

When the simulation begins, the defects repel each other and move apart, and they rotate toward the orientation favored by the boundary conditions.  After that, the behavior depends on the sign of the activity $Z_0$.  If $Z_0<0$, as in Fig.~4(a-d), the defects move until they reach the edge of the active region.  At that point, they are confined by the activity gradient, and cannot leave the active region.  As a result, they circulate around the active region, in the direction parallel to their $\bm{p}$ vectors (as expected for negative, contractile activity).  They remain in this state of circular motion indefinitely.  By contrast, if $Z_0>0$, as in Fig.~4(e-h), the defects move directly through the edge of the active region.  Once they are in the inactive region, they are no longer driven by activity, and hence they stop moving.

We have also done simulations (not shown here) with the radial boundary condition $\theta_{BC}(x,y)=\tan^{-1}(y/x)$, which favors an \emph{outward} alignment of the $+1/2$ defect orientation vectors.  In that case, the dependence on the sign of $Z_0$ is reversed:  If $Z_0>0$, the defects circulate indefinitely around the active region, in the direction opposite to their $\bm{p}$ vectors (as expected for positive, extensile activity).  If $Z_0<0$, the defects move directly through the edge of the active region, and stop moving in the inactive region.

These simulation results can be understood based on the theory presented earlier in this paper.  In Fig.~4(a-d), the boundary conditions favor an inward defect orientation, while the activity gradient favors an outward defect orientation.  Hence, the defects are oriented inward by the boundaries, but then they have an unfavorable interaction with the activity gradient around the edge of the pink circle.  The edge of the circle is effectively a wall on the defects, which confines them within the circular region.  For that reason, they move in a circle inside that region.  In Fig.~4(e-h), the boundary conditions and the activity gradient both favor an inward defect orientation.  Hence, the defects rotate to an inward orientation, and they can easily move through the activity gradient.  The edge of the pink circle is not a wall for them, and they go directly into the inactive region.  If the boundary condition is reversed from tangential to radial, then the dependence on the sign of $Z_0$ is also reversed.  Thus, these simulation results provide one example of how an activity gradient can be designed to guide the motion of topological defects.

In summary, this paper has presented three approaches to model the interaction of topological defects with nonuniform activity:  hydrodynamic theory based on the director and velocity fields, macroscopic theory based on symmetry, and simulations based on the nematic order tensor and velocity fields.  All of these approaches show that an activity gradient aligns the orientation vector of a $+1/2$ defect, in a similar way to an electric field aligning an electric dipole moment.  These results agree with the previous work of other groups using different methods~\cite{Shankar2019,Zhang2019}, and support the concept that nonuniform activity patterns can be used to control active materials.

\acknowledgments

This work was supported by National Science Foundation Grant DMR-1409658.

\bibliography{activitygradient4}

\end{document}